# Quantum-Classical Reentrant Relaxation Crossover in $Dy_2Ti_2O_7$ Spin-Ice

J. Snyder[1], B. G. Ueland[1], J. S. Slusky[2], H. Karunadasa[2], R. J. Cava[2], Ari Mizel[1], and P. Schiffer[1*]

[1]*Department of Physics and Materials Research Institute, Pennsylvania State University, University Park PA 16802*

[2]*Department of Chemistry and Princeton Materials Institute, Princeton University, Princeton, NJ 08540*

## ABSTRACT

We have studied spin relaxation in the spin ice compound $Dy_2Ti_2O_7$ through measurements of the a.c. magnetic susceptibility. While the characteristic spin relaxation time is thermally activated at high temperatures, it becomes almost temperature independent below $T_{cross}$ ~ 13 K, suggesting that quantum tunneling dominates the relaxational process below that temperature. As the low-entropy spin ice state develops below $T_{ice}$ ~ 4 K, the spin relaxation time increases sharply with decreasing temperature, suggesting the emergence of a collective degree of freedom for which thermal relaxation processes again become important as the spins become strongly correlated.

[*]schiffer@phys.psu.edu

Geometrically frustrated magnetic materials, in which the geometry of the spin lattice leads to frustration of the spin-spin interactions, have been shown to display a wide range of novel ground states [1,2]. There has been especially strong recent interest in the so-called "spin ice" pyrochlore materials (such as $Dy_2Ti_2O_7$, $Ho_2Ti_2O_7$, and $Ho_2Sn_2O_7$) in which the rare-earth spins are highly uniaxial due to strong crystal field effects [3]. Ferromagnetic and dipolar interactions between these spins on this lattice of corner-sharing tetrahedra are frustrated in a manner analogous to that of protons in ice leading to a variety of exotic behavior [4,5,6,7,8,9,10,11,12,13,14,15,16,17,18,19]. While the spin entropy only freezes out below $T_{ice}$ ~ 4 K in $Dy_2Ti_2O_7$ [12], magnetic susceptibility studies show a strongly frequency dependent spin-freezing at $T$ ~ 16 K [13,14]. In contrast to traditional spin glasses, the spin-freezing transition is associated with a very narrow range of relaxation times and thus represents a rather unusual example of glassiness in a dense magnetic system.

Here we report a study of the spin relaxation processes in the spin ice compound $Dy_2Ti_2O_7$. We find that, while the characteristic spin relaxation time is thermally activated at high temperatures, it becomes almost temperature independent below $T_{cross}$ ~ 13 K. We interpret the data in terms of a crossover from thermal to quantum mechanical relaxation mechanisms as a function of temperature. While most previous studies of such thermal-quantum crossovers involve isolated moments, this material represents a unique situation of quantum relaxation in a dense spin system with developing correlations. These correlations result in a highly unusual re-emergence of thermally activated behavior for $T \lesssim T_{ice}$ as the collective degrees of freedom dominate the spin dynamics at low temperatures.

Polycrystalline $Dy_2Ti_2O_7$ samples were prepared using standard solid-state synthesis techniques described previously [13]. X-ray diffraction demonstrated the samples to be single-phase, and Curie-Weiss fits done to the high temperature susceptibility were consistent with $J = 15/2$ $Dy^{3+}$ ions. We study the magnetization ($M$) and the resultant d.c. susceptibility ($c_{dc} = dM/dH$) as well as the real and imaginary parts ($c'$ and $c''$) of the a.c. susceptibility ($c_{ac}$). The d.c. susceptibility measurements were performed with a Quantum Design MPMS SQUID magnetometer, and the a.c. susceptibility measurements were made with the ACMS option of the Quantum Design PPMS cryostat or with a simple inductance coil in a dilution refrigerator at low temperatures.

In figure 1 we show the temperature dependent susceptibility at different frequencies. We find that $c_{dc}$ increases monotonically with decreasing temperature as expected for a paramagnetic system with no spin-freezing. While $c'(T)$ is virtually identical to $c_{dc}$ at our lowest frequency, at higher frequencies $c'(T)$ has a sharp decrease at $T_f \sim 16$ K, deviating well below $c_{dc}$ [13,14]. This sharp drop leads to a local maximum in $c'(T)$ correlated with a sharp rise in $c''(T)$ at a "freezing" temperature, $T_f$, which increases with frequency. This feature is a common signature of a glass transition, implying that the system is out of equilibrium on the time scale of the measurement. While magnetic site dilution studies [13] indicated that this freezing was associated with the development of spin-spin correlations, recent data suggest that it may instead be a single-ion effect [18]. As demonstrated in the inset to figure 1, the frequency dependence of $T_f$ can be fit to an Arrhenius law, $f = f_0 e^{-E_a/k_B T_f}$ where $f_o$ ( $\sim 10^9$ Hz) is a measure of the microscopic attempt frequency in the system and $E_A$ is an activation energy for fluctuations ($E_A \sim$ 200-300 K), which is the energy scale set by the crystalline field [3,13,14].

The Arrhenius law behavior indicates that the spin relaxation processes are thermally driven for $T \gtrsim T_f$. The frequency dependence of the freezing is extraordinarily strong, $\Delta T_f / T_f \Delta(\log f) \sim 0.18$, an order of magnitude higher than is typical for spin glasses [20]. An extrapolation of this strong frequency dependence suggests that one should observe $T_f \sim 8$ K for frequencies of order 10 mHz, the timescale of the nominally static magnetization measurements. The absence of observed irreversibility in the magnetization in this temperature range implies that the characteristic relaxation time does not continue to follow the Arrhenius law to lower temperature.

The spectrum of spin relaxation times can be determined from the frequency dependence of $\chi''$, which we have measured as a function of temperature and field (figure 2). For a single characteristic relaxation time, $\tau$, $\chi''(f)$ has a maximum at $f = 1/\tau$ and is given by the Lorentzian form $\chi''(f) = f\tau \left( \dfrac{\chi_T - \chi_S}{1 + f^2 \tau^2} \right)$ where $\chi_T$ is the isothermal susceptibility in the limit of low frequency and $\chi_S$ is the adiabatic susceptibility in the limit of high frequency [21]. Experimentally one typically finds broadening associated with a spread of relaxation times, but $\chi''(f)$ for $Dy_2Ti_2O_7$ in zero magnetic field is only slightly broader than the single-$\tau$ form for our entire temperature range [13], indicating the existence of a temperature dependent characteristic spin relaxation time with a narrow distribution at any given temperature. The spectrum does broaden somewhat in a magnetic field, but there is a clear local maximum to $\chi''(f)$ for almost our entire temperature and magnetic field range, which we take to indicate the characteristic relaxation time, $t$, where $t = 1/f_{max}$. While the form of the peak in $c''(f)$ is somewhat broadened relative to the ideal case and there is a broad background underneath the peak at the lowest temperatures and highest fields, the resultant uncertainty in the determination of the peak frequency does not qualitatively change the temperature or magnetic field dependence of $t$, and thus does not affect the discussion below.

As shown in figure 3, the temperature dependence of $t$ for $T > T_{ice}$ is quite striking. Above a crossover temperature, $T_{cross} \sim 13$ K, $t$ is thermally activated, as would be expected from the Arrhenius behavior of $T_f$. Below $T_{cross}$, however, we find that $t$ is almost temperature independent ($t \sim 5$ ms at H = 0) down to $T \sim T_{ice}$. This reduced temperature dependence explains the absence of spin freezing in the d.c. magnetization data, since the time scale of that measurement is still much longer than $t$. The weak temperature dependence raises a new question of how the spins are relaxing for $T < T_{cross}$, since $t(T)$ is not consistent with a thermally activated process (which would require an unphysically weak energy barrier of $E_A < 2K$ for transitions between the two spin states). Since the Hamiltonian of this pure ordered system is relatively simple, it is difficult indeed to imagine how any combination of thermal processes could result in the observed $t(T)$. We conclude instead that the relaxation process for $T_{ice} < T < T_{cross}$ is through quantum tunneling between the two accessible Ising states, as has been suggested for other systems in which the relaxation rate changes from being thermally activated to almost constant [22] and inferred for $Ho_2Ti_2O_7$ from higher and lower temperature neutron spin echo studies [18].

As shown in figures 3 and 4, the relaxation time has a non-monotonic dependence on applied d.c. magnetic field [23]. For small fields (H < 5 kOe), $t$ decreases with increasing field, but for larger fields $t(H)$ increases and then saturates (the field required for saturation increases with decreasing temperature). We interpret the measured $t(H)$ in terms of a simplified spin Hamiltonian given by: $H = -DS_z^2 - g\mathbf{m}_B \vec{S} \cdot \vec{H}_{applied} - g\mathbf{m}_B \vec{S} \cdot \vec{H}_{dipole}$ where $D$ represents the local anisotropy, $\vec{H}_{applied}$ is the applied d.c. field, $\vec{H}_{dipole}$ is the local dipolar field, and we neglect the relatively weak exchange interactions [5]. Given the four different [111] directions of the Dy spins in $Dy_2Ti_2O_7$ and that our samples are polycrystalline, the applied field has components both transverse and along the axis of almost all spins in the sample. The low field decrease in $t(H)$ can be partially attributed to enhanced quantum tunneling through coupling to the transverse components of $\vec{H}_{applied}$ [24]. The applied field also can enhance tunneling by locally canceling the longitudinal component of $\vec{H}_{dipole}$, and thus making the $\pm S_z$ states degenerate. This latter effect is evident in a numerical evaluation [25] of the quantum tunneling barrier from the above Hamiltonian (averaging over the directions of both $\vec{H}_{dipole}$ and $\vec{H}_{applied}$), which indicates that $t(H)$ should exhibit a minimum when the cancellation is optimized. The corresponding minimum in the data of figure 4 shifts to lower field at higher temperatures, as expected from the reduction of the mean dipolar field with increasing thermal fluctuations.

The increasing $t(H)$ at higher magnetic fields is attributable to the longitudinal component of $\vec{H}_{applied}$ separating the energies of the $\pm S_z$ states. The weaker field dependence of $t(H)$ at the highest fields (at 12 K, for example) presumably marks the return to thermal relaxation processes when the $\pm S_z$ states are highly separated by $\vec{H}_{applied}$. A difference between the physics of tunneling at low and high fields is suggested by the magnitude of the maxima in $c''(f)$, shown in the inset to figure 4, which increase monotonically with decreasing temperature in zero field (as expected from the reduced dissipation at lower temperatures), but decrease monotonically at 10 kOe. These data suggest strong coupling of phonons to the quantum spin-relaxation at the higher fields. The presence of such coupling, previously observed in Mn-12 acetate [22], would account for the energy released or absorbed when a spin flips to be either aligned or anti-aligned with the d.c. field. The rich behavior of $t(H,T)$ in $Dy_2Ti_2O_7$ suggests that detailed modeling including the phonon spectrum will be necessary for a complete description of the quantum tunneling phenomena.

While a crossover between thermal and quantum spin relaxation has been observed in many systems, the case of $Dy_2Ti_2O_7$ is particularly interesting because it is a dense system in which spin-spin correlations are developing as $T \rightarrow T_{ice}$. Below $T_{ice}$, a maximum in $c'(T)$ and irreversibility in the d.c. magnetization [14] suggest that $\tau$ again increases with decreasing temperature as the spin ice correlations develop at low temperatures. We have also performed direct measurements of $t(T)$ for $T < T_{ice}$ in a powder sample of $Dy_2Ti_2O_7$ potted in epoxy for thermal contact (figure 3 inset) [26]. Surprisingly, while $\tau(T)$ is almost temperature independent for $T_{ice} < T < T_{cross}$, it increases extremely sharply below $T_{ice}$ (faster than Arrhenius law behavior). This low temperature increase in $\tau(T)$ indicates the existence of an unusual double-crossover with decreasing temperature. The relaxation process changes from thermally activated to quantum tunneling at $T_{cross}$, and then reverts to thermally activated as the spin-correlations become sufficiently strong to require that groups of spins must respond coherently to changes in the applied field. The authors are unaware of similar behavior in other systems exhibiting spin relaxation through quantum tunneling, suggesting that this phenomenon is associated with the unusual development of correlations in the frustrated spin ice state and the associated emergence of a collective degree of freedom of the sort which has been previously observed in frustrated magnets [27].

Aside from the novelty of a double crossover in the relaxation mechanism, the observation of quantum relaxation in $Dy_2Ti_2O_7$ indicates the importance of such processes in frustrated rare-earth magnets, which have been shown to display a range of novel low temperature states including spin liquids and spin glasses in the presence of minimal disorder [28]. These results imply that even such large spin systems, where the spins are typically treated classically, must be considered with the effects of quantum dynamics. As the emergence of thermal relaxation of the correlated spins at low temperatures indicates, the combination of quantum and thermally activated processes in a strongly correlated spin system can form the basis for unanticipated physical phenomena associated with the collective properties of such spin systems.

**ACKNOWLEDGEMENTS**

We gratefully acknowledge helpful discussions with D. Huse, and R. Moessner and support from the Army Research Office PECASE grant DAAD19-01-1-0021. R. J. C. was partially supported by NSF grant DMR-9725979. A. M. gratefully acknowledges the support of a Packard Foundation Fellowship.

**Figure Captions**

Figure 1. Temperature dependence of the magnetic susceptibility of $Dy_2Ti_2O_7$ in the absence of a d.c. magnetic field. **a**. The d.c. susceptibility and real part of the a.c. susceptibility ($c'$). Inset: the d.c. magnetization versus applied field. **b**. The imaginary part of the a.c. susceptibility ($c''$). Inset: the measurement frequency versus inverse freezing temperature in zero applied field, fit to an Arrhenius law between 10 Hz and 10 kHz.

Figure 2. The frequency dependence of the imaginary part of the a.c. susceptibility ($c''$) in various applied fields at $T$ = 7.5, 12, and 16 K. The bar in the 16K plot shows the full width at half maximum (FWHM) of the theoretical response of a system with a single relaxation time.

Figure 3. The characteristic relaxation time, $t$, as a function of temperature at various applied fields as determined by the position of the local maximum of $c''(f)$. The inset shows the low temperature behavior of $t(T)$ at $H = 0$ for a sample potted in epoxy [26].

Figure 4. The magnetic field dependence of the characteristic relaxation time, $t$, as determined by the position of the local maximum of $c''(f)$. The inset shows the magnitude of the peak in $c''(f)$ as a function of temperature at $H = 0$ and 10 kOe.

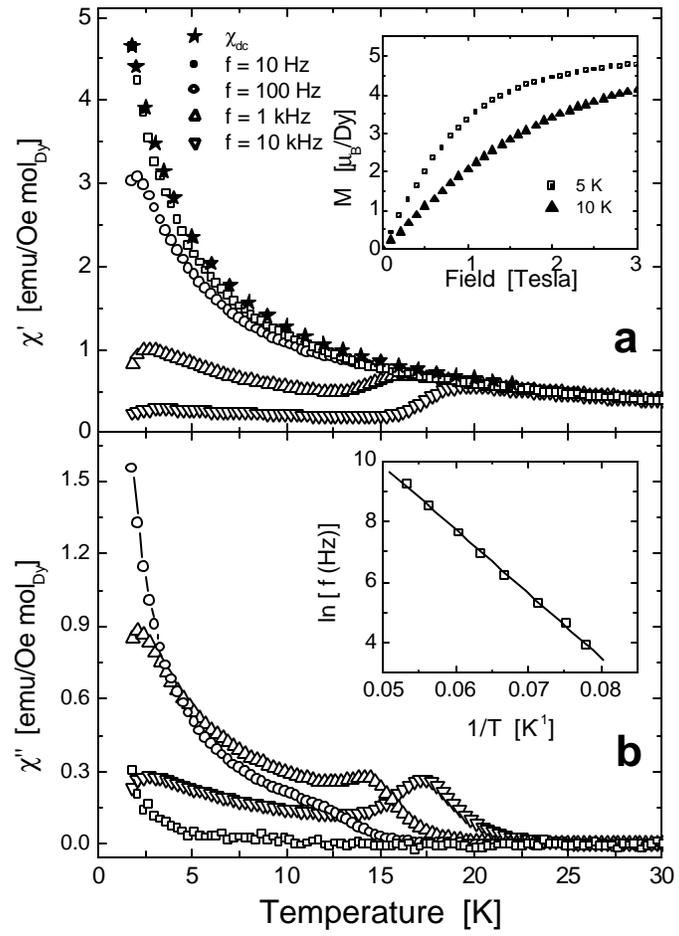

Figure 1 Snyder *et al*.

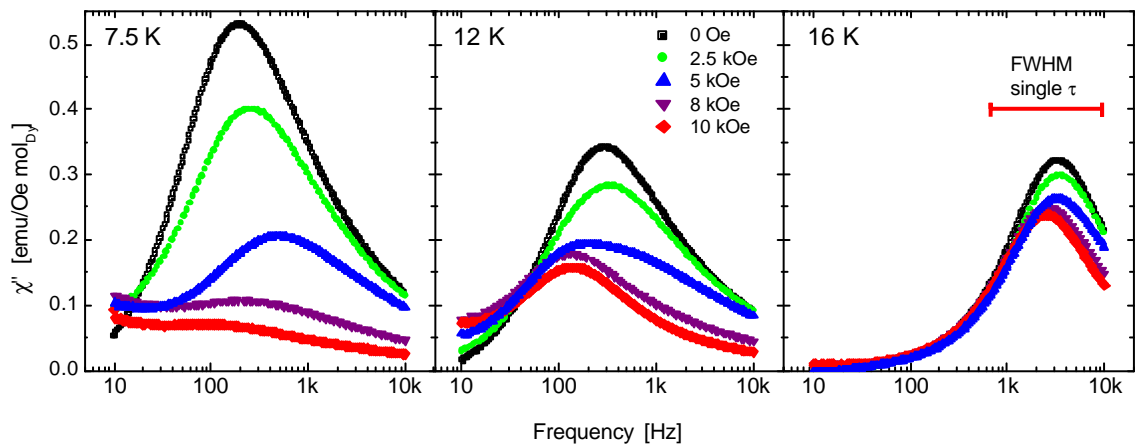

Figure 2  Snyder *et al.*

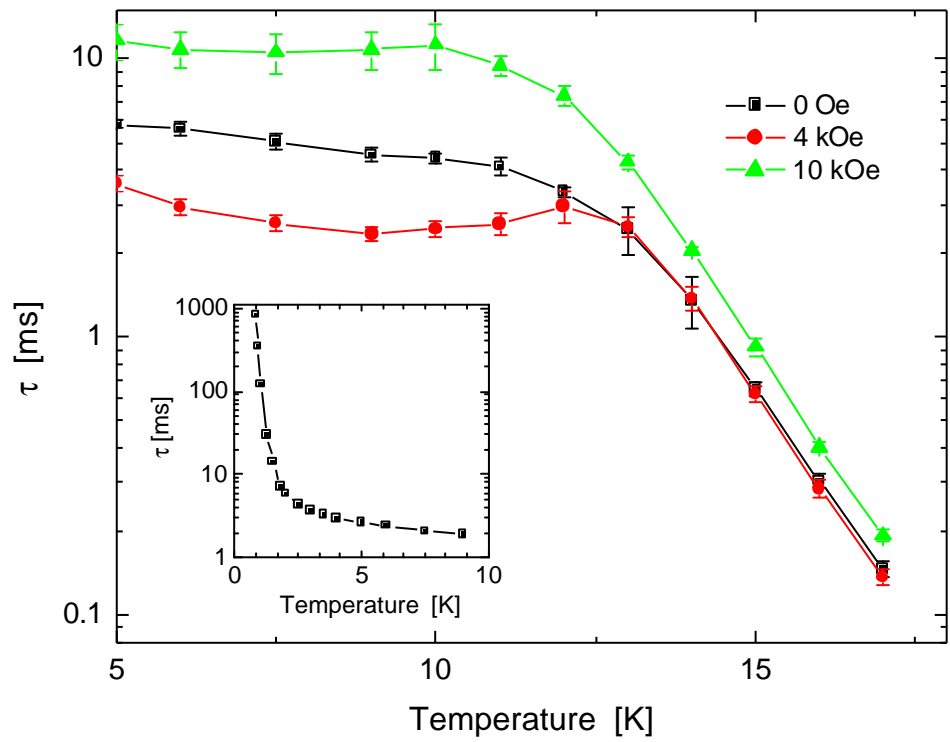

Figure 3 Snyder *et al.*

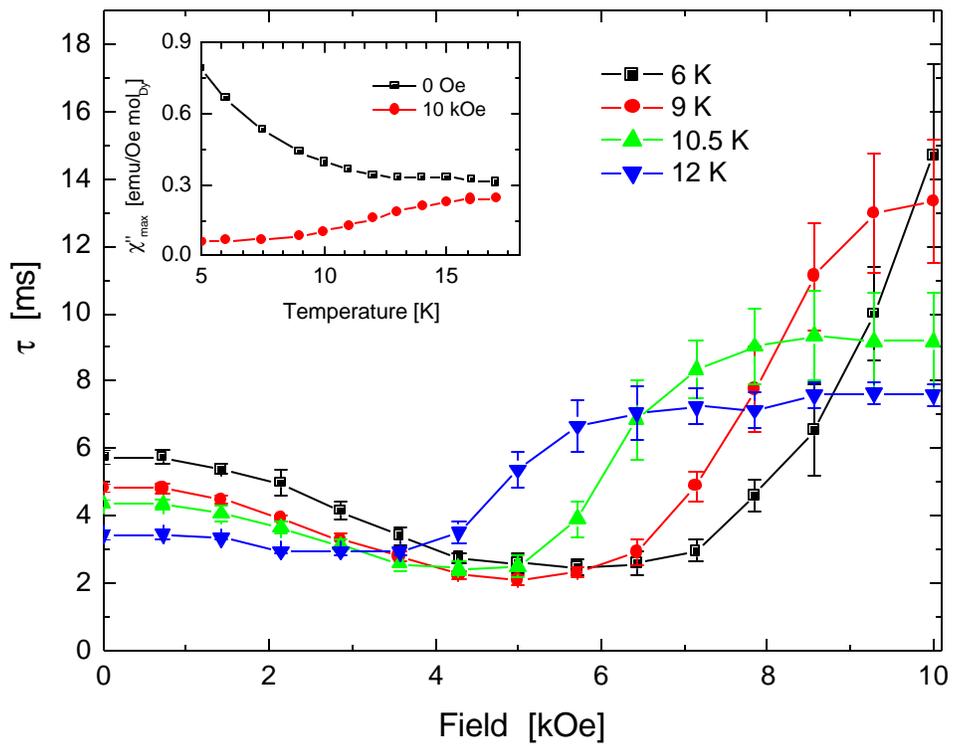

Figure 4  Snyder *et al.*